# Magnetic Breakdown and Topology in the Kagome Superconductor $CsV_3Sb_5$ under High Magnetic Field

Ramakanta Chapai,[1,*] Maxime Leroux,[2] Vincent Oliviero,[2] David Vignolles,[2] M. P. Smylie,[1,3]
D. Y. Chung,[1] M. G. Kanatzidis,[1,4] W.-K. Kwok,[1] J. F. Mitchell,[1] Ulrich Welp[1,†]
[1]*Materials Science Division, Argonne National Laboratory, Lemont, IL 60439, USA*
[2]*LNCMI-EMFL, CNRS UPR3228, Univ. Grenoble Alpes, Univ. Toulouse, INSA-T, Grenoble and Toulouse, France*
[3]*Department of Physics and Astronomy, Hofstra University, Hempstead, NY 11549, USA*
[4]*Department of Chemistry, Northwestern University, Evanston, IL 60201, USA*

The recently discovered layered Kagome metals of composition $AV_3Sb_5$ ($A$ = K, Rb, Cs) exhibit a complex interplay among superconductivity, charge density wave order, topologically non-trivial electronic band structure and geometrical frustration. Here, we probe the electronic band structure underlying these exotic correlated electronic states in $CsV_3Sb_5$ with quantum oscillation measurements in pulsed fields up to 86 T. The high-field data reveal a sequence of magnetic breakdown orbits that allows the construction of a model for the folded Fermi surface of $CsV_3Sb_5$. The dominant features are large triangular Fermi surface sheets that cover almost half of the folded Brillouin zone that have not yet been detected in angle resolved photoemission spectroscopy (ARPES). These sheets display pronounced nesting at the charge density wave (CDW) vectors, which may stabilize the CDW state. The Berry phases of the electron orbits have been deduced from Landau level fan diagrams near the quantum limit without the need for extrapolations, thereby unambiguously establishing the non-trivial topological character of several electron bands in this Kagome lattice superconductor.

*rchapai@anl.gov; †welp@anl.gov

The interplay of superconductivity, charge density wave (CDW) order and non-trivial topology of the electronic band structure in the metallic Kagome lattice compounds $AV_3Sb_5$ ($A$ = K, Rb, and Cs) has generated substantial interest in the effects arising from van Hove singularities, Dirac crossings, electronic correlations, and geometrical frustration [1-7]. For instance, an unusual relationship between the CDW and superconductivity (SC) has been reported via pressure and doping studies: mechanical or chemical pressure induces a double-dome shaped superconducting phase diagram in $CsV_3Sb_5$ [8-12]. If the SC and CDW states compete for the same electrons at the Fermi surface [13, 14], suppressing CDW order should enhance $T_c$, suggesting an unconventional SC state in $CsV_3Sb_5$. The CDW order in $CsV_3Sb_5$ is itself fascinating due to the presence of chiral charge order and time-reversal symmetry breaking without local magnetic moments [15-19]. This state is believed to give rise to a large anomalous Hall effect [4, 5, 20]. The transition into the CDW state is accompanied by an in-plane unit cell doubling [3] and the concomitant folding of the Fermi surface and appearance of CDW-induced gaps which mostly affect bands derived from vanadium *d*-orbitals. Furthermore, in the out-of-plane direction, doubling [21, 22] or quadrupling [3] of the unit cell have been reported.

Detailed knowledge of the reconstructed electronic band structure is prerequisite for understanding the interplay between different charge orders, flux phases and possible unconventional superconductivity in the $AV_3Sb_5$ family. While angle resolved photoemission spectroscopy (ARPES) has been invaluable in exploring the electronic structure, effects due to matrix-elements have largely precluded the visualization of the folded band-structure [23, 24]. In contrast, quantum oscillations (QOs) are a direct manifestation of the actual Fermi surface revealing information on the quasiparticle effective masses, their lifetimes and their topological state. Here, we report on quantum oscillations observed in high-quality single crystal $CsV_3Sb_5$ in pulsed magnetic fields up to 86 T. Twenty-five distinct frequencies ranging from 20 T to 9930 T have been identified, with most of the high-frequency bands as new findings. In particular, the high-field data yield a sequence of magnetic breakdown orbits that allow us to construct a model for the folded Fermi surface of $CsV_3Sb_5$. The dominant features are large triangular Fermi surface sheets that cover almost half of the folded Brillouin zone (BZ) that have not been seen via ARPES [6, 24]. These sheets display pronounced nesting at the CDW vectors which may stabilize the CDW state. Using the Lifshitz-Kosevich (LK) formalism [25] we determine effective masses ranging from $0.17m_0$ to $2.02m_0$. The Berry phases of the electron orbits are deduced from Landau level (LL) fan diagrams. Due to the very high magnetic field, we can evaluate the phase of quantum oscillations near the quantum limit, thereby unambiguously revealing the non-trivial topological character of the $\beta$ ($F_{\beta}$ = 79 T), the $\xi_1$ ($F_{\xi 1}$ = 736 T), and the $\xi_2$ ($F_{\xi 2}$ = 804 T) bands. The latter two bands are derived from Dirac crossings at the H and K points in the BZ, respectively, which our results place at 290 meV and 350 meV below the Fermi level.

Quantum oscillation measurements using the tunnel diode oscillator (TDO) technique [26-28] were performed at the pulsed-field facility in the Toulouse National Intense Magnetic Field Laboratory (LNCMI-T) CNRS. Figure 1(a) displays the field dependence of the TDO frequency ($f_{TDO}$) measured in magnetic fields applied parallel to the *c*-axis.

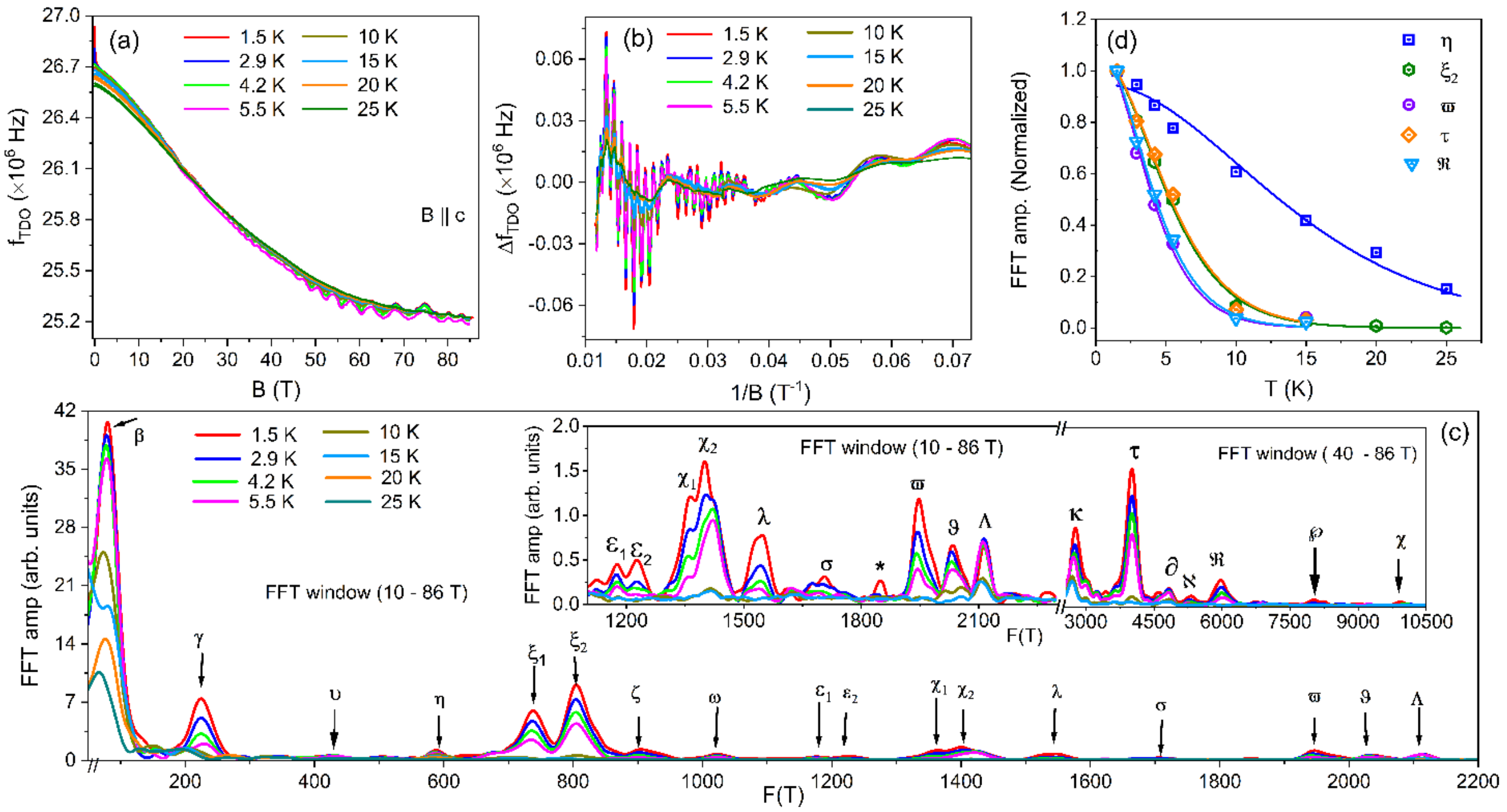

Figure 1. (a) Field dependence of the TDO frequency ($f_{TDO}$) of $CsV_3Sb_5$ at various temperatures. (b) SdH oscillations, visible in $\Delta f_{TDO}$ after the subtraction of the non-oscillatory part, plotted as function of $1/B$. (c) FFT spectrum of data presented in frame (b). Inset: the FFT spectra displaying high frequencies. The numbers in parenthesis denote the field range used for FFT analysis. (d) Temperature dependence of the FFT amplitudes of several bands. Solid lines are fits to the thermal damping term of Eq. (S1) [29].

The quantum oscillation observed as the oscillatory part of the TDO frequency, $\Delta f_{TDO}$, obtained after subtracting a smooth background [29] is plotted as a function of inverse field ($1/B$) in Fig. 1(b). A complex spectrum with 25 distinct oscillation frequencies emerges from the fast Fourier transformation (FFT) analysis of the data with frequencies ranging from 20 T to 9930 T as shown in Fig. 1(c). The very high frequencies exceeding 2700 T are new findings reported here. Table S1[29] summarizes all the frequencies shown in Fig. 1(c).

As the oscillatory TDO signal is a measure of the temperature and field dependence of the conductivity, we use the LK formalism for the analysis of Shubnikov-de Haas (SdH) oscillations [25, 30, 31], see Supplemental Material [29] for more details. Figure 1(d) displays the temperature dependence of the FFT amplitudes for several bands. From fits of the observed temperature dependences to the thermal damping factor $R_T$ (Eq. (S1)[29]) we obtain cyclotron effective masses $m_{\eta}^{*} = 0.19 m_0$, $m_{\xi 2}^{*} = 0.52 m_0$, $m_{\varpi}^{*} = 0.69 m_0$, $m_{\tau}^{*} = 1.57\ m_0$, and $m_{\Re}^{*} = 2.02 m_0$. Data for additional bands are shown in Fig. S3 and in Table S1 [29]. In addition to bands with very light masses, already seen in previous studies [3, 32-36], we observe 'heavier' bands with effective masses approaching 2. Such heavier bands are expected as van Hove singularities are contributing to the electronic structure near the Fermi energy [6, 37].

The 1st BZ of a hexagonal lattice with lattice constant $a$ has an area of $8\pi^2/(\sqrt{3}\ a^2)$ which, in the case of $CsV_3Sb_5$, is ~1.55 Å$^{-2}$ corresponding to a quantum oscillation frequency of 16.24 kT. In the CDW state, the BZ is folded to ¼ of its pristine size, that is, a frequency of about 4060 T. This implies that all the high frequencies observed in Fig. 1(c) either represent harmonics of lower frequency orbits or are caused by magnetic breakdown or quantum interference [25]. Magnetic breakdown [25, 38-41] can arise when two extremal orbits on different sheets of the Fermi surface are so close (in $k$-space) that an electron (hole) can tunnel from one orbit to the other and continue its trajectory thereby inducing sum and difference frequencies in the spectrum. Their collective behavior can be described in a model of coupled networks of semiclassical electron trajectories [38, 42] yielding a relation for the oscillatory energy that is similar to the LK formula albeit with an additional damping factor characterizing the tunneling events. The prevalence of magnetic breakdown is indicated by Dingle plots [25, 43] displaying a clear non-monotonic behavior (Fig. S4 [29]).

The spectrum in Fig. 1(c) reveals a series of high-frequency orbits, $F_{\varpi}$ = 1943 T, $F_{\tau}$ = 4025 T, $F_{\Re}$ = 5996 T, $F_{\mathfrak{L}}$ = 8014 T, $F_{\mathfrak{X}}$ = 9930 T that are approximately equally spaced at ~2000 T. Such a sequence could arise due to higher harmonics of the $\varpi$-orbit, magnetic breakdown orbits or quantum interference involving multiple copies of the $\varpi$-orbit. Higher harmonics have effective cyclotron masses that are enhanced by the harmonic order over that of the fundamental, whereas the effective mass of breakdown orbits (between like carriers) and quantum interferences equals the sum and difference of the involved masses, respectively [41, 44, 45]. Experimentally, we find that the effective masses of the high-frequency orbits $F_{\varpi}$, $F_{\tau}$ and $F_{\mathcal{R}}$ scale approximately as multiples (see Table S1[29] and above) thus ruling quantum interference out as the mechanism of high-frequency orbits. Our data indicate that the successive frequencies correspond to the addition of a unit to the orbit that is robust in very high fields and that occupies close to half of the folded BZ. While the details of the reconstructed Fermi surface are currently not well established [34], the high-field behavior yields additional constraints and allows construction of a model that captures the observed behavior, as shown in Fig. 2 (see also Fig. S5 [29]). The schematic (Fig. 2(a)) suggests that the triangular orbit (shown in magenta) centered on the H-point and spanned by three A-points labeled 1, 2, 4 is the most likely building block for the high-frequency orbits (see Fig. 2(b)), and we identify it with $F_{\varpi}$.

The experimental frequencies are systematically larger than expected for a harmonic, $n\times F_{\varpi}$. The $\tau$-orbit has a frequency of 4025 T which is very close to the full folded BZ, labeled 1, 2, 3, 4 in Fig. 2(a), see also Fig. S6 (a) [29]. This orbit can arise through magnetic breakdown near two A-points (2, 4) as depicted in Fig. 2(a), implying that this orbit has an area of two triangular $\varpi$-orbits plus the area of the narrow lamella centered on L between two close magenta sheets, the area of which we estimate to be ~100 T, even though the details of the FS near the L-point are uncertain. The 120-degree turns in the $\tau$-orbit at locations 2 and 4 trace the orbit of the pristine FS and are therefore natural to arise during breakdown (see Fig. 2(c)). Following this scheme, we identify the $\mathcal{R}$-orbit (5996 T) with three triangular plaquettes (see Fig. S6 (b)), the $\mathcal{L}$-orbit (8014 T) with four and a line at $F_{\mathfrak{X}}$ = 9930 T with five triangular plaquettes. Thus, our high-field observations reveal a remarkable network of reconstructed Fermi surface sheets that displays a high degree of nesting at the CDW vectors. This property may serve to stabilize the 3-Q charge ordered state [7]. As shown in Fig. 2(a), the Fermi velocity is essentially uni-directional over extended sections of the Fermi surface which is expected to yield characteristic signatures in magneto-transport [46] and thermodynamic properties such as the temperature dependence of the upper critical field or of the penetration depth. DFT calculations indicate that the $\varpi$-band persists at the H, K and L points whereas near the M-point it is gapped [6, 7, 33]. Detailed angular dependent studies in high fields will be required to establish the three-dimensional structure of the reconstructed Fermi surface.

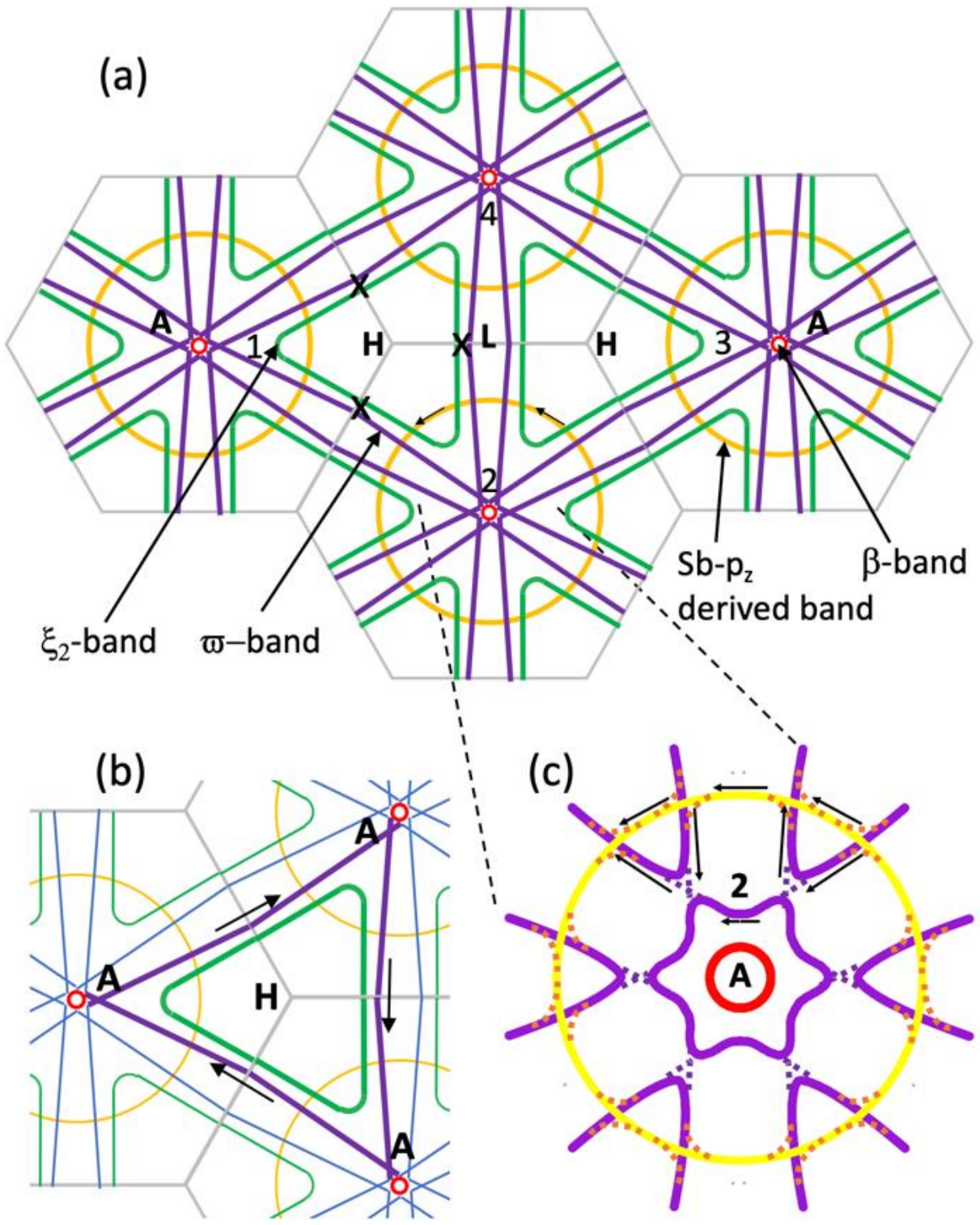

Figure 2 (a): Schematic of the 2x2 folded Fermi surface of $CsV_3Sb_5$ for $k_z = \pm\pi/c$ in repeated zones. The crosses indicate likely breakdown junctions between the $\xi_2$ and $\varpi$ orbits. (b) Enlarged schematic highlighting the $\xi_2$ and $\varpi$ orbits centered on the H-point. (c) Enlarged schematic showing details around the A-point as indicated by the dashed lines. For clarity, $\xi_2$-orbits are not included, and the yellow sheet is drawn at reduced size as compared to (a). The locations of breakdown are indicated as dashed lines. The arrows indicate the semi-classical trajectories of the carriers.

Inspection of Fig. 2(a) also suggests that the $\kappa$-orbit (2753 T) results from the breakdown between the $\varpi$-orbit and the triangular orbit centered on the H-point (presented in green, Fig. 2(b)). Based on recent quantum oscillation measurements and band-structure calculations [33], we identify this triangular sheet with the $\xi_2$ orbit having a frequency of 804 T. Accordingly, $\xi_1$ (736 T) represents the corresponding triangular sheet centered on the K-point ($k_z$=0). A carrier can tunnel from the $\varpi$-orbit to the $\xi_2$-orbit at the point of smallest separation on the LH-line marked by crosses in Fig. 2(a), complete the $\xi_2$-orbit, tunnel back to ϖ at the same point and complete the ϖ-orbit resulting in a sum frequency of $F_{\varpi}$+$F_{\xi 2}$ of 2743 T which is close to the observed frequency of $F_{\kappa}$ = 2753 T. Furthermore, the measured effective mass of $1.13m_0$ is close to the expected value of $1.21m_0$. Similarly, the $\partial$-orbit with a measured frequency of $F_{\partial} = 4827$ T results from the same breakdown mechanism between the $\tau$ and the $\xi_2$-orbits with an expected frequency of 4829 T. In addition to a complete $\xi_2$-orbit, the carrier can tunnel back to the ϖ-orbit at the next point of close proximity after completing 1/3 of the $\xi_2$ orbit or at the following point of close proximity after completing 2/3 of the $\xi_2$ orbit. These trajectories will generate quantum oscillation frequencies of 2/3 $F_{\varpi}$ + 1/3 $F_{\xi 2}$ = 1547 T and of 1/3 $F_{\varpi}$ + 2/3 $F_{\xi 2}$ = 1172 T, respectively, in very good agreement with the observed frequencies $F_{\lambda}$ = 1550 T and $F_{\varepsilon 1}$ = 1170 T. The corresponding measured effective masses of $0.86m_0$ and $0.54m_0$ appear in reasonable agreement with the expected values of $0.66m_0$ and $0.57m_0$. The $F_{\beta}$~79 T frequency, accompanied by a small effective mass, has been observed consistently in all previous works [3, 4, 32-36]. These parameters are close to those reported [33] for a small prolate ellipsoid shaped FS near the L point in the pristine BZ of $CsV_3Sb_5$ which, upon reconstruction, appears at the A-point of the folded zone (small red circle in Fig. 2). The star-shaped orbit near the zone center arises from intersecting and gapping of the magenta sheets, see Fig. 2(c). We identify this orbit with $F_{\gamma}$ motivated by its small size and the fact that it has the same effective mass as the $\varpi$-orbit. This orbit is an electron orbit. Our data did not enable the identification of the circular electron orbits centered on Γ−A and derived from Sb $p_z$-states (shown in yellow in Fig. 2). A likely cause is the large number of breakdown junctions, 24, encountered on these orbits making them difficult to observe in quantum oscillations. Furthermore, some bands in Fig. 1(c) remain un-assigned, most notably ϑ, Λ, and the $\chi_1$-$\chi_2$ pair which has the same splitting, 68 T, as $\xi_1$ and $\xi_2$. Possibly, these frequencies are associated with the 3D-structure of the Fermi surface, orbits in the $k_z$ = 0 plane and $c$-axis reconstruction.

Following the LK formalism, the phase of the oscillatory conductivity at a given frequency reveals the Berry phase, that is, the topological state of the orbit in question. The oscillatory component of $\Delta f_{TDO}$ for a given frequency is isolated via a filtering process [47]. Figure 3(a) shows $\Delta f_{TDO}$ versus $B^{-1}$ for the β-band isolated by using a band-pass filter (60-100 T). The Landau fan diagram is constructed by assigning the SdH oscillation minima to $N$, the LL index, and maxima to $N$ +1/2 [30]. As shown in Fig. 3(b) for the β-orbit, $N(B^{-1})$ can be fitted as $N = 0.40 + F/B$ yielding a frequency of $F_{\beta}$ = 85.43 T, in good agreement with that obtained from the FFT spectra (Fig. 1(c)). This implies that the filtering process to isolate the β-orbit preserves the original signal.

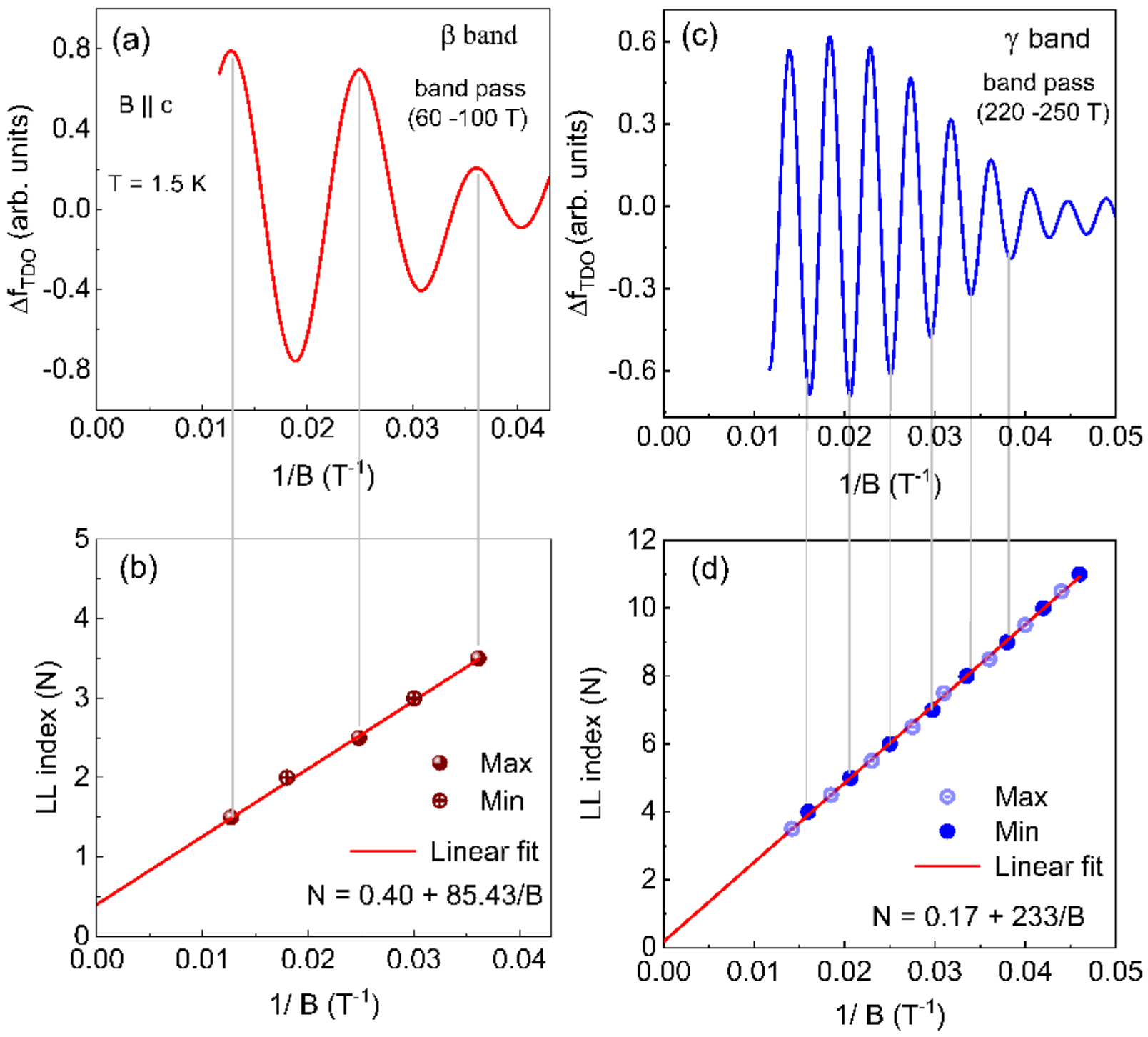

Figure 3. (a) $\Delta f_{TDO}$ of the β band plotted as a function of inverse magnetic field ($B^{-1}$). (b) Landau fan diagram constructed from the SdH oscillations in frame (a). (c)-(d) Similar data for the γ-band. The solid lines in frames (b) and (d) are linear fits of the data.

The intercept of this linear equation gives the Berry phase as $(\Phi_B^{\beta}/2\pi) + \delta = 0.40$. This band has previously been identified as a prolate electron ellipsoid [33], thus δ = -1/8 [48]. With this consideration, $\Phi_B^{\beta} \sim \pi$, a non-trivial Berry phase [49]. As the charge carriers reach the first LL (near the quantum limit) at 86 T for the β band, we can determine the intercept in the LL plot without extrapolation from high LL indices thus establishing the non-trivial topology of the β band. Similarly, the LL fan diagrams constructed for the $\gamma$-orbit (Fig. 3(c)-(d)) and $\eta$-orbit (Fig. S7 (a) [29]) yield intercepts of ~0.2 and ~0, respectively. Thus, these two orbits have trivial topology. Even though the very high frequency bands ($\tau$ and $\mathscr{R}$ bands) can be isolated with filtering, hence enabling the construction of the LL fan diagrams, determining the Berry phase for those bands requires an extrapolation from high LL indices (Fig. S7 (g-h) [29]). In addition, magnetic breakdown can induce additional phase shifts [25, 38-40] which induces uncertainty in the determination of the Berry phase. As demonstrated in Fig. S7 (d-h) [29], the LL fan diagrams for the $\omega$, $\kappa$-, $\tau$- and $\mathscr{R}$-orbits display the intercepts of ~0, ~0.5, ~0 and 0.5 respectively and the linear fits yield frequencies in good agreement with the spectral analysis. Furthermore, we verified that the extrapolations do not depend on the field range (Fig. S7 [29]) demonstrating that there are no field-induced changes in the electron-orbits in high fields. For the $\xi_1$ and $\xi_2$ bands, we obtain Berry phases of ~1.32π and 1.14π (shown in Fig. S7 (c, d) [29]) identifying these bands as non-trivial, consistent with them being derived from Dirac crossings at the H and K points, respectively. Using the linear relativistic Dirac dispersion, given as $E - E_D = \hbar v_D k$, and an area of the orbit, assumed for these estimates to be circular, we obtain the location of the Dirac crossing, $E_F$-$E_D$, and the Dirac velocity, $v_D$, in terms of measured quantities as $v_D = \sqrt{2e\hbar F}/m^*$ and $E_F - E_D = 2e\hbar F/m^*$, yielding $v$ ~ $2.9\times10^5$ m/s and $3.5\times10^5$ m/s, and $E_F$-$E_D$ ~ 287 meV and 353 meV for the $\xi_1$ and $\xi_2$ bands, respectively. These estimates for the location of the Dirac crossings are in good agreement with band structure calculations and ARPES results [6, 7, 33, 50, 51].

In summary, we discovered sequences of high field breakdown orbits on the reconstructed Fermi surface of $CsV_3Sb_5$ in pulsed magnetic fields up to 86 T. These results enable the construction a consistent model for the folded Fermi surface of $CsV_3Sb_5$. The dominant features are large triangular Fermi surface sheets that cover almost half of the folded Brillouin zone and are absent in ARPES measurements. A series of QO frequencies can be understood as arising from breakdown orbits encompassing multiples of the triangular sheets. Using the LK formalism, we determined effective masses ranging from $0.17m_0$ to $2.02m_0$, and from LL fan diagrams, deduced the Berry phases of the electron orbits. The very high magnetic field enabled us to evaluate the phase of quantum oscillations near the quantum limit without the need for extrapolations, thereby unambiguously revealing the non-trivial topological character of the β ($F_\beta$ = 79 T) band, and of the $\xi_1$ ($F_{\xi1}$ = 736 T) and $\xi_2$ ($F_{\xi2}$ = 804 T) bands. The latter two bands are derived from Dirac crossings at the H and K points, respectively. A reconstructed Fermi surface containing extended almost flat sections should yield characteristic signatures in magneto-transport and thermodynamic properties such as the temperature dependence of the upper critical field or of the penetration depth.

## Acknowledgments

This work was supported by the U. S. Department of Energy, Office of Science, Basic Energy Sciences, Materials Sciences and Engineering Division. We acknowledge the support of the LNCMI-CNRS, member of the European Magnetic Field Laboratory (EMFL), for the TDO measurements at Toulouse National Intense Magnetic Field Laboratory (LNCMI-T). We thank A. E. Koshelev and M. R. Norman for helpful discussions.

# Supplemental Material

## Experimental details

Single crystals of $CsV_3Sb_5$ were grown using the flux method [1-3]. $Cs_3Sb_7$ (0.1 g), $VSb_2$ (0.024 g), and Sb (0.05 g) were mixed in an alumina crucible and sealed in a fused silica tube. The mixture was then heated at 1000 $^oC$ for 24 h, cooled to 650 $^oC$ at a rate of 2 $^oC/h$, and subsequently quenched to room temperature. As-grown single crystals were structurally characterized through x-ray diffraction (XRD) measurement in a PANAlytical X'Pert Pro diffractometer with Cu $K_\alpha$ radiation. Electrical resistivity measurements were performed using a dc-technique in a Physical Properties Measurement System (DynaCool-PPMS, Quantum Design) following a standard four-probe method with current applied along the in-plane direction. Quantum oscillation measurements using the TDO technique [4-6] were performed at the pulsed-field facility in the Toulouse National Intense Magnetic Field Laboratory (LNCMI-T) CNRS. Single crystal samples with typical dimensions of 100 x 100 x 50 $\mu m^3$ were placed on a spiral-like flat copper coil that is part of the TDO circuit, resonating at ~25 MHz, and that is mounted in a $^4He$ cryostat. A high-speed data acquisition system was used to digitalize the signal. The data were post-analyzed to extract the field dependence of the resonance frequency which is sensitive to the conductivity of the sample through the change in the sample's skin depth. Here, we present data on the down-sweep which has lower dB/dt and therefore affords better thermalization of the sample.

## Sample Characterization

Figure S1(a) shows the XRD pattern of $CsV_3Sb_5$ single crystal indexed with the hexagonal structure (space group *P*6/*mmm*). An optical image of a typical $CsV_3Sb_5$ single crystal is shown in the inset of Fig. 1(a) revealing the overall hexagonal morphology. The largest flat surface corresponds to the *ab* basal plane, confirmed by XRD showing a preferred orientation along the [00*l*] direction. The lattice parameter *c* is calculated to be 9.304(5) Å, a value close to that reported previously [3, 7]. The temperature dependence of the in-plane electrical resistivity $\rho_{ab}(T)$ shown in Fig. S1(b) displays two anomalies, one at $T_c$ ~3.2 K corresponding to SC and another at $T_{CDW}$ ~94 K arising due to CDW. These findings are consistent with previously reported values [2, 3]. The samples display a low residual resistivity ~1.2 μΩ cm at ~ 4 K before entering the superconducting state reflecting their high-quality.

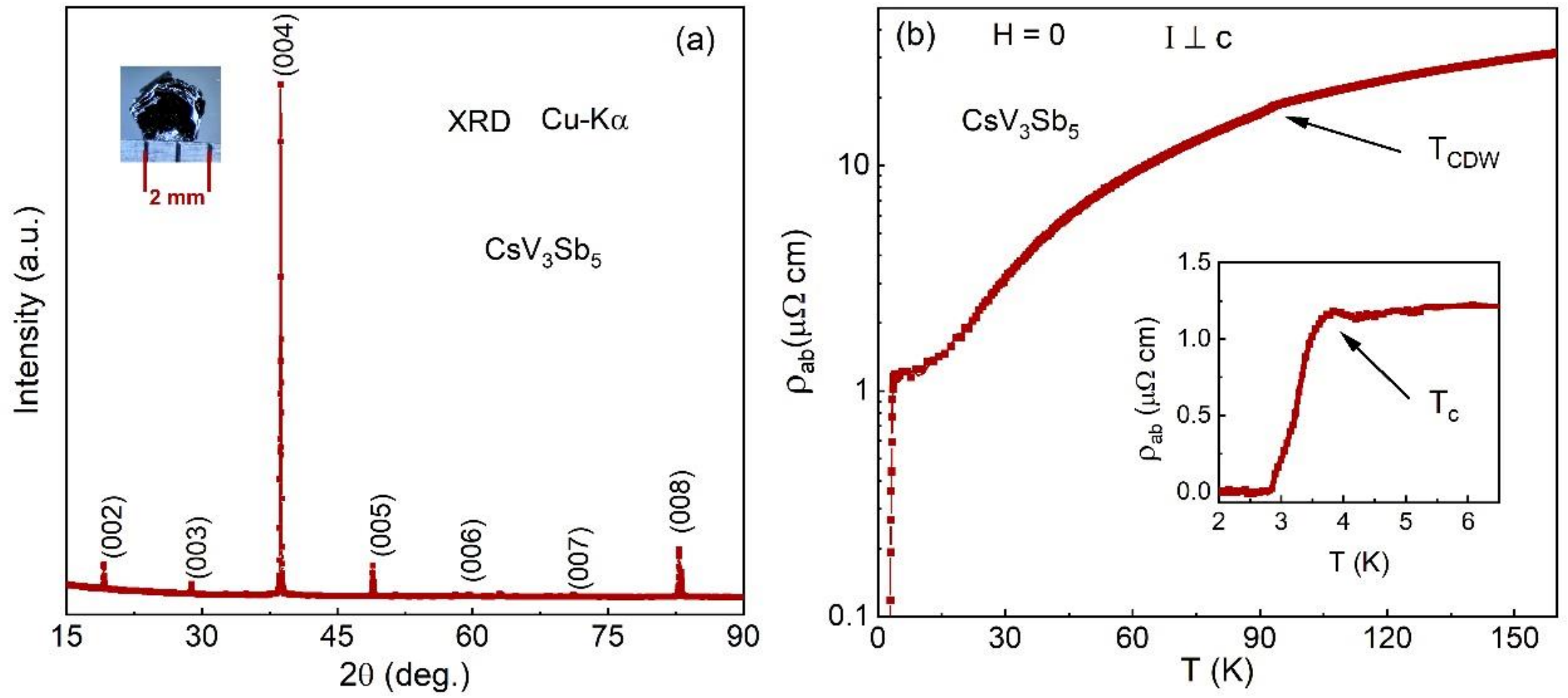

Figure S1. (a) The room temperature XRD pattern of $CsV_3Sb_5$ single crystal indexed with the *P*6/*mmm* structure. Inset: An optical image of a typical $CsV_3Sb_5$ single crystal. (b) Temperature dependence of the electrical resistivity $\rho_{ab}$ (*T*) of $CsV_3Sb_5$. Inset: $\rho_{ab}$ (*T*) versus *T* at low temperature displaying superconducting transition with $T_c$~3.5 K.

### Detail of background subtraction

The experimental data (blue, Fig. S2(a)) are fitted with a third order polynomial (red line, Fig. S2(a)) which is then subtracted from the experimental data. The same procedure is applied to each data set at different temperature to account for any temperature dependent background signal. Besides the SdH oscillations in the high field regime (Fig. S2(b)), signatures of superconductivity at low fields can also be seen in the background subtracted data at temperatures below $T_c$ (~3.5 K).

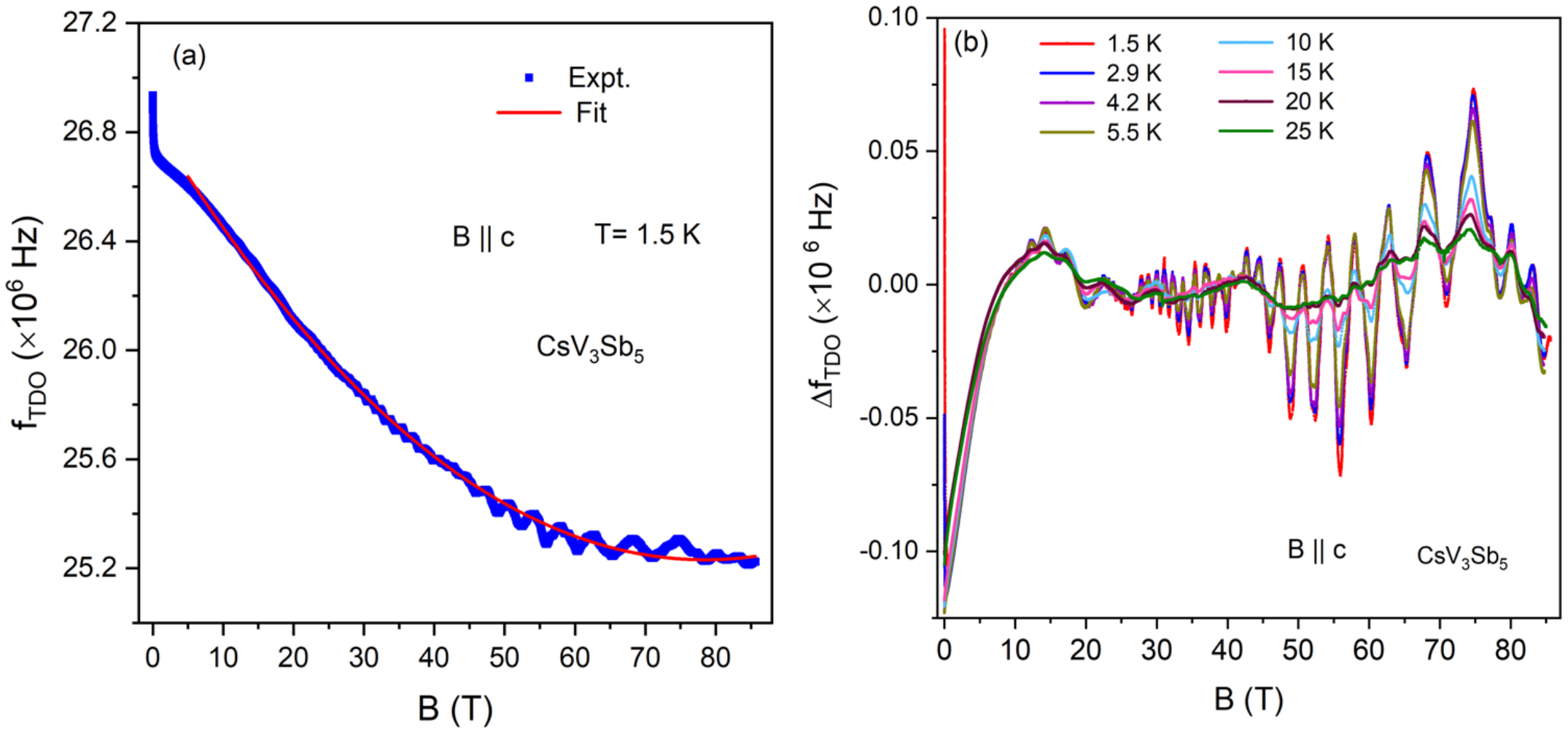

Figure S2: (a) Field dependence of the TDO frequency at 1.5 K. The red curve is a third order polynomial fit to the data. (b) Field dependence of the TDO frequency at indicated temperatures after subtracting the third order polynomial as non-oscillatory background.

**Lifshitz-Kosevich formalism**

The oscillatory part of the conductivity is given as [8-10]:

$$\Delta\sigma_{xx} = A_o \left(\frac{B}{F}\right)^{\lambda} R_T R_D R_S \cos\left[2\pi \left\{\frac{F}{B} - \left(\frac{1}{2} - \Phi\right)\right\}\right] \quad \text{(S1)}$$

where $F$ (in units of Tesla) is the frequency of the oscillation, $R_T = \frac{A\left(\frac{m^*}{m_0}\right)\frac{T}{\langle B\rangle}}{Sinh\left(A\left(\frac{m^*}{m_0}\right)\frac{T}{\langle B\rangle}\right)}$ is the thermal damping factor, $A = \frac{2\pi^2 k_B m_0}{e\hbar}$ = 14.69 T/K, $R_D = exp\left(-A\frac{m^*}{m_0}\frac{T_D}{B}\right)$ is the Dingle damping factor ($T_D$ is the Dingle temperature), $R_S = \cos(\pi g^* \frac{m^*}{2m_0})$ is the spin reduction factor, $m^*$ is the thermal cyclotron effective mass given as $m^* = \hbar^2/2\pi\ \partial S_k/\partial E|_{E_F}$, and $g^*$ is the effective $g$ factor. The effective field, $1/\langle B\rangle$, entering the thermal damping factor is determined through the lower and upper limits of the FFT window used in determining the oscillation amplitudes as $1/\langle B\rangle = (1/B_{min} + 1/B_{max})/2$. The exponent λ is equal to 0 for a 2D system and ½ for a 3D system. In addition, $\Phi = \frac{\Phi_B}{2\pi} + \delta$ where $\Phi_B$ is the Berry phase and δ = 0 for a 2D and $\pm 1/8$ for a 3D FS, where $\pm$ sign corresponds to minima (+) / maxima (-) of the cross-sectional area of the FS for the case of an electron band. For a 3D hole band, the sign of δ is opposite [11]. We restrict the analysis of the field and temperature dependence of the oscillation amplitudes to orbits that are sufficiently well separated in the spectrum.

## Mass Plots

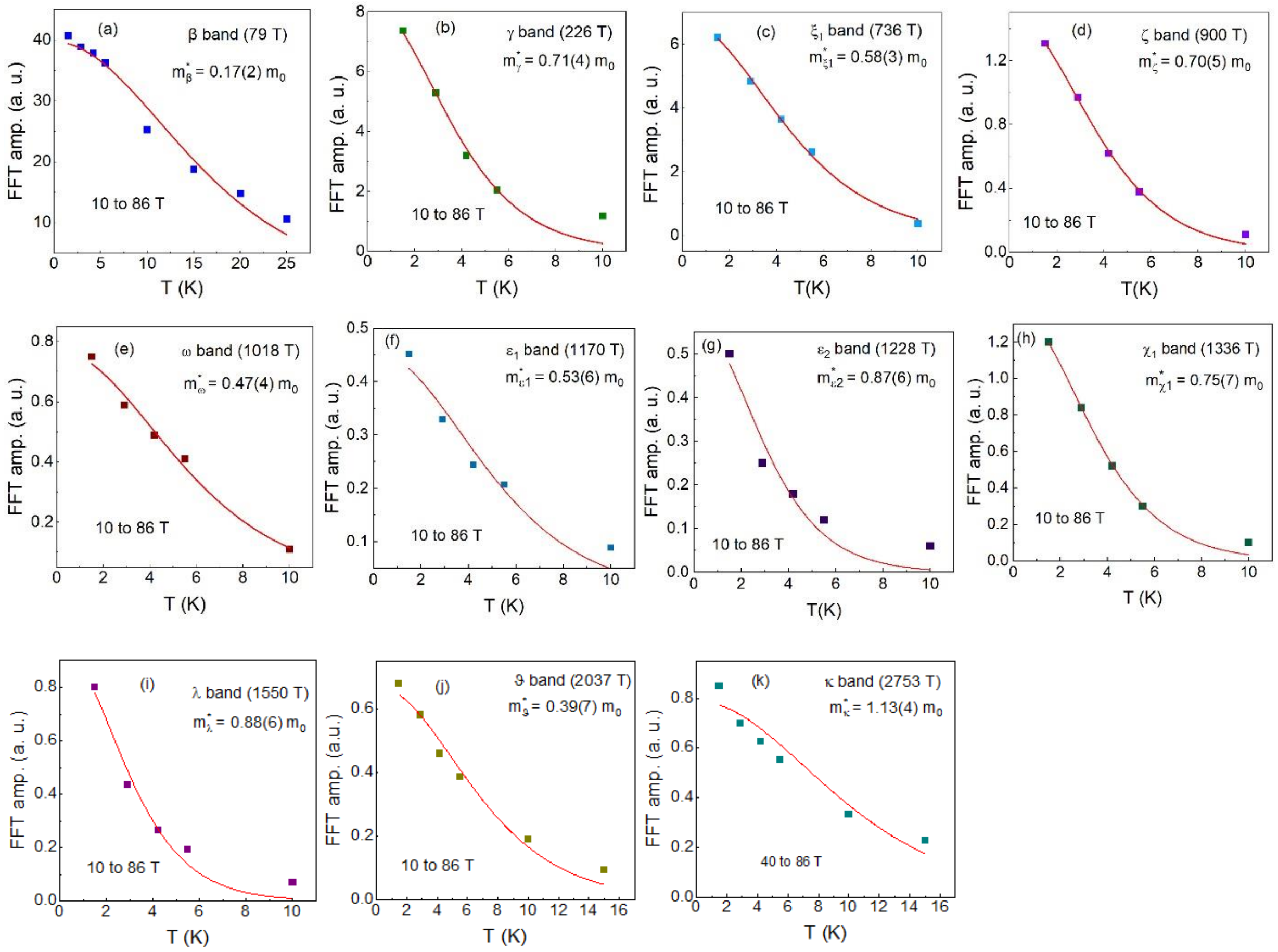

Figure S3: Temperature dependence of the FFT amplitude of various bands shown in Fig. 1(c) (main text).

## Dingle Plots

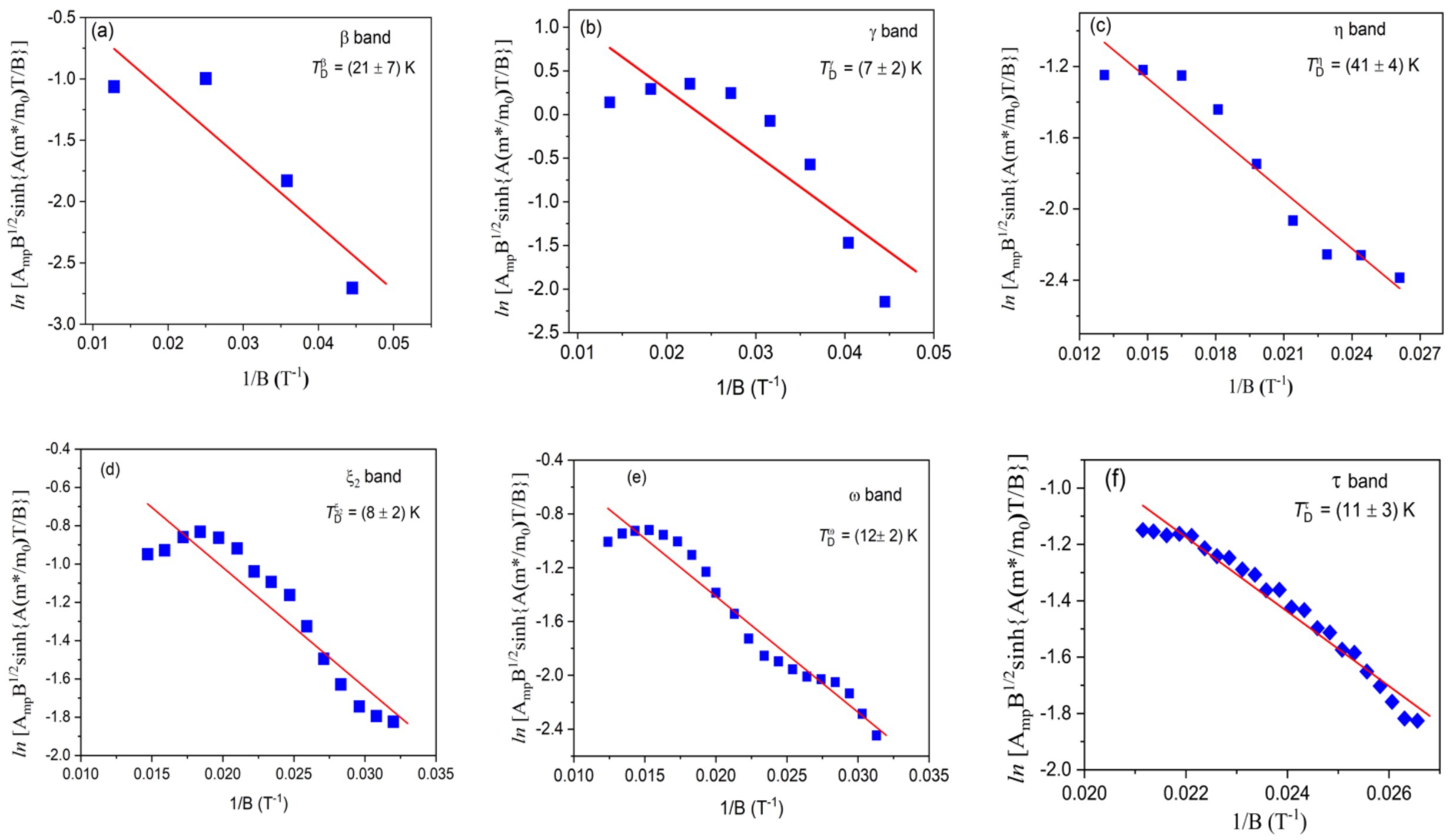

Figure S4: Dingle plots of (a) η band and (b) τ band with $B \parallel c$. Linear fits are included to give an overall estimate of the magnitude of $T_D$.

The Dingle plots for various orbits reveal a clear downwards curvature and/or non-monotonic behavior. While these behaviors are generally expected for magnetic breakdown [8], a quantitative fit and a determination of the magnetic breakdown field $B_0$ could not be obtained since $T_D$ and $B_0$ enter the LK-expression in very similar ways preventing an independent determination of either [12].

**Folded Brillouin Zone**

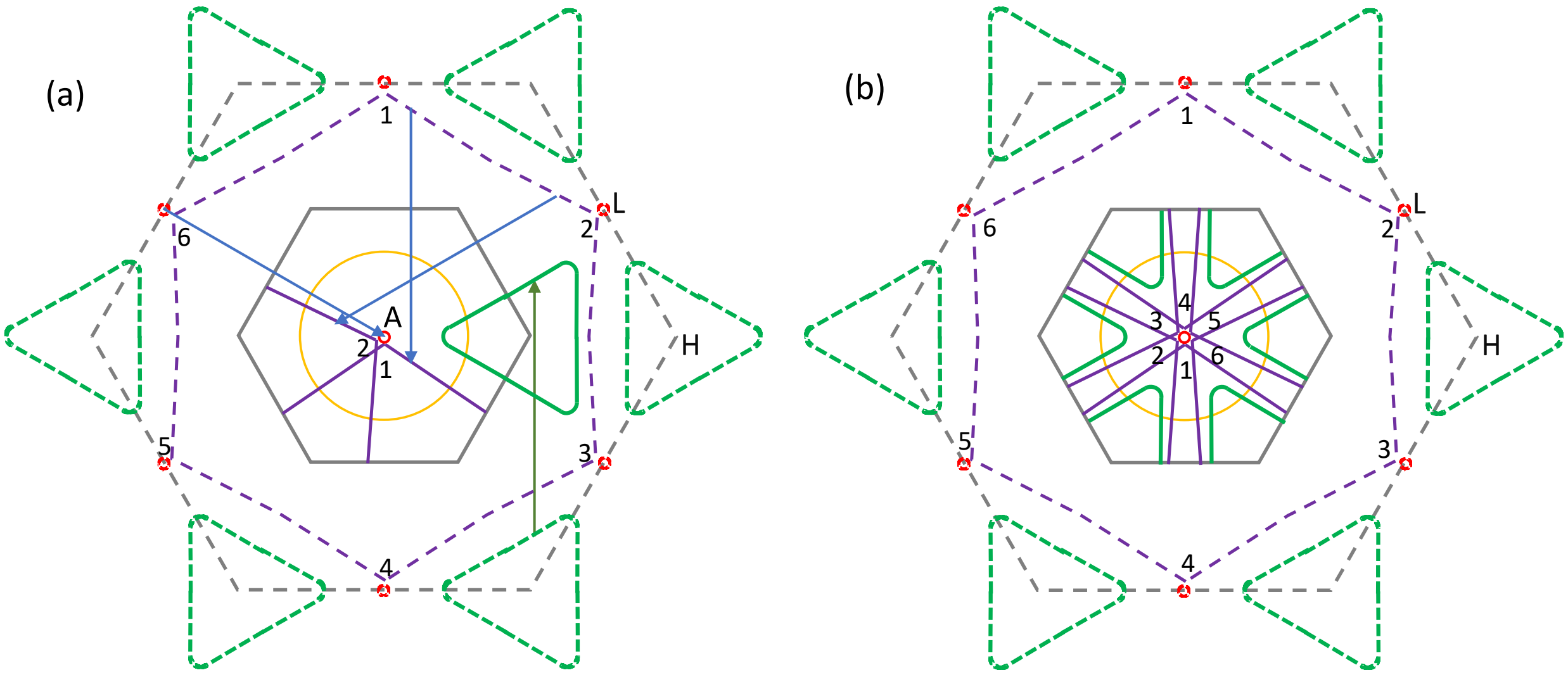

Figure S5: Construction of the folded Fermi surface in the $k_z = \pm\pi/c$ plane. (a) Dashed and solid grey lines mark the BZ in the pristine and in the CDW state, respectively. Correspondingly, the dashed and solid-colored lines represent Fermi surface sheets in the pristine and CDW states. The arrows represent the propagation vectors of the CDW. Numbers 1 through 6 mark the locations near the L-points of the pristine BZ and the corresponding locations in the folded zone, respectively. The yellow circle in the center of the zone represents an electron sheet derived from Sb-$p_z$ states. Panel (a) highlights how Fermi surface sections around points "1" and "2" are transformed into the folded zone. Completion of this procedure for locations "1" through "6" yields the propeller-shaped Fermi surface sheet shown in magenta in panel (b) and Fig. 2 [main text]. (b) the folded BZ in relation to the original zone.

Figure S5 shows the construction of the folded Fermi surface. The 1st Brillouin zone (BZ) in the pristine and in the CDW state are shown as dashed and solid hexagons, respectively. The arrows in panel (a) indicate the propagation vectors of the CDW. As the CDW is commensurate with the pristine structure, the folded FS results from translating FS sheets by the CDW vectors as shown in panel (a). In particular, the dashed magenta line suspended by numbers 1 through 6 represents a large hole surface in the pristine BZ which upon folding yields the propeller-shaped intersecting pattern shown in Fig. 2 of the main text in repeated zones and in panel (b) in the first zone. The detailed structure of the Fermi surface surrounding the H (K) points depends sensitively on the location of the Fermi energy with respect to a band gap near the H (K) points in the BZ [3, 13-15]. For instance, ARPES results [15, 16] indicate a double-$k_F$ crossing consisting of nearly overlapping electron and hole sheets at the Fermi energy located on the Γ(A) – K(H) lines. Here, we represent the electronic structure around H(K) with a single sheet (dashed and solid green triangles) with a size modelled after [3, 15]. This sheet may have a convex [13, 15] or a concave [3] shape with rounded corners. We note that an increased size of this sheet would cause overlap and intersection with the magenta sheets in the folded zone. Our QO results indicate that such overlap does not occur. Here, we do not consider reconstruction along the *c*-axis.

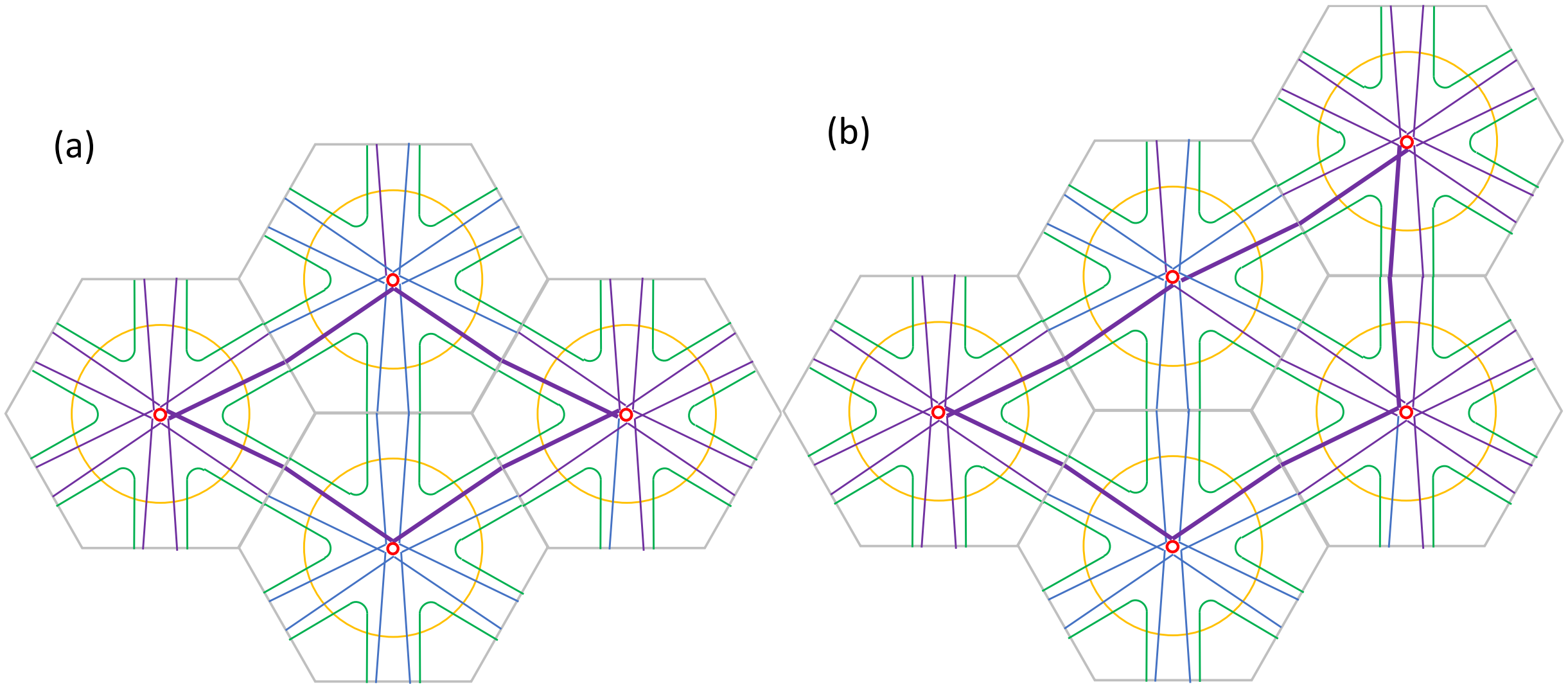

Fig. S6: Depiction of (a) the τ-orbit composed of two plaquettes having an area corresponding to almost the entire 1st BZ and (b) of the $\mathcal{R}$-orbit containing three plaquettes.

## Landau Level fan diagrams

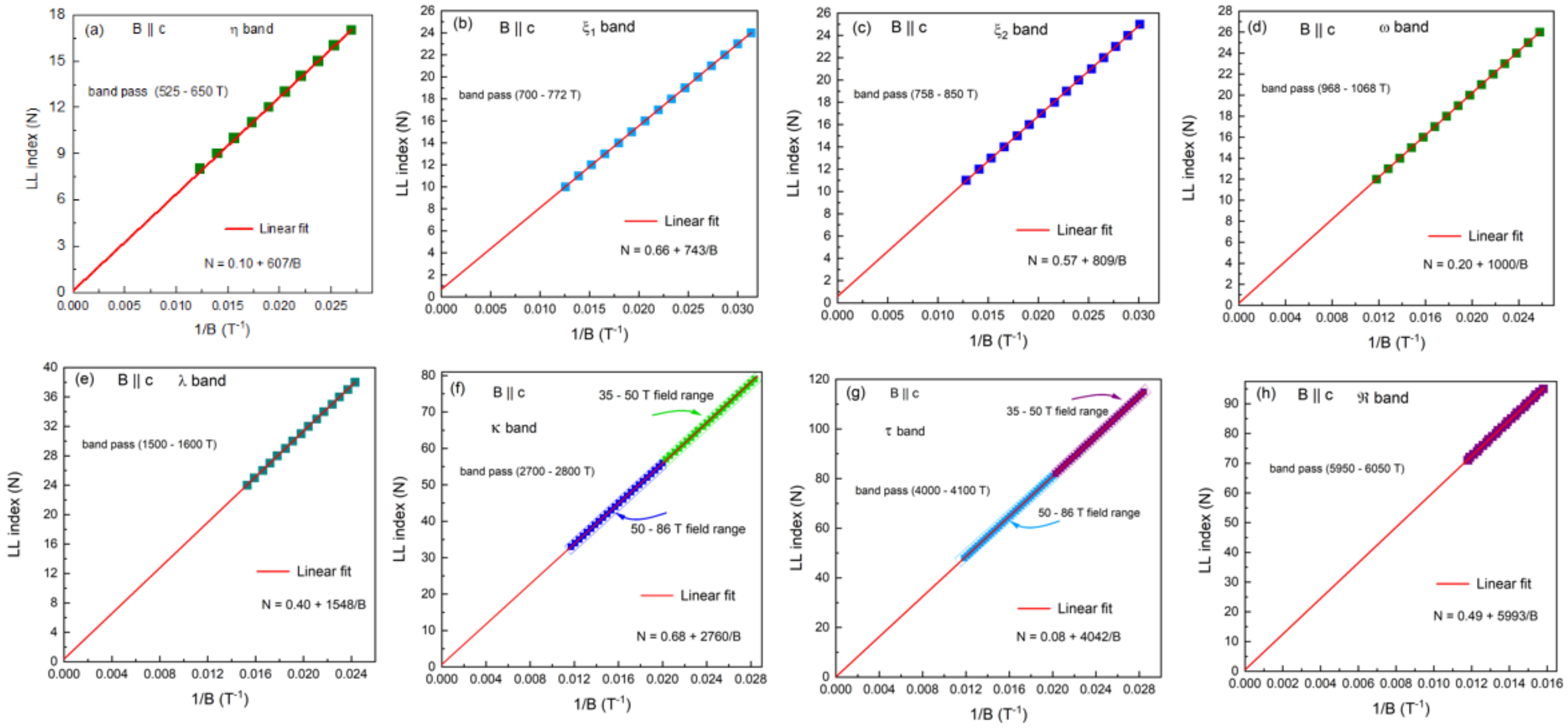

Figure S7: Landau level fan diagrams of various bands. The band pass filters employing to isolate different bands are included in the respective frames. For high frequency bands ($\kappa$ and $\tau$), different field range considered to construct the LL fan diagram is also indicated in the respective frames.

Table S1: List of frequency *F* (in unit of Tesla) that have been detected from quantum oscillations (dHvA & SdH) in $CsV_3Sb_5$ with $B \parallel c$ listed with the maximum field used in the measurement in the parenthesis. The number in the parenthesis is the effective mass in units of $m_0$. NTr denotes a nontrivial and Tr denotes a trivial Berry phase. The frequencies are classified as low, medium, high and ultra-high so as to compare with the existing literature. TrH denotes the case of trihexagonal distortion.

| Bands | F (This work, 86 T) | F (Ref. [2], 35 T) | F (Ref. [1], 14 T) | F (Ref. [3], 32 T) | F (Ref. [7], 14 T) | F (Ref. [17], 14 T) | F (Ref. [18], 14 T) | F (Ref. [19], 9 T) | F (DFT) |
|---|---|---|---|---|---|---|---|---|---|
| Low frequency ($F_l$) | | | | | | | | | |
| — | — | — | 11 | — | 18 (0.028) | 18 (0.06) | 12 | 18 (0.039) | With ISD distortion, a prolate ellipsoid of 73 T around L point [1,3] |
| α | 20 | 18 | 28 | 27 | 26 (0.031) | 28 (0.07) | 27 | 28 (0.043) | |
| β | 79 (0.17, NTr) | 102 (0.346, NTr) | 74(0.09) /90 (0.11) | 73 (0.142, NTr) | 72/92 | 72(0.11) /88(0.12) | 73 | 72(0.058) /91(0.054) | |
| γ | 226 (0.71, Tr) | 239 (0.327, NTr) | — | — | — | — | — | — | |
| υ | 429 | — | — | — | — | — | — | — | |
| η | 588 (0.19, Tr) | — | 580 | — | — | — | | | |
| $\xi_1$ | 736 (0.58, NTr) | 788 (0.302, NTr) | — | 727 (0.54, NTr) | — | — | — | — | A round triangular orbit of ~800 T at K point [3] |
| $\xi_2$ | 804 (0.52, NTr) | 865 (0.312, NTr) | 860 | 786 (0.55, Tr) | — | — | — | — | |
| Medium frequency ($F_m$) | | | | | | | | | |
| ζ | 900 (0.71) | — | — | — | — | — | — | — | Circular FS of 1123 T at Γ (Sb p-orbital [1]) |
| ω | 1018 (0.50, Tr) | — | — | — | — | — | — | — | |
| $\varepsilon_1$ | 1170 | — | — | — | — | — | — | — | |
| | (0.54) | | | | | | | | |
| $\varepsilon_2$ | 1228 (0.88) | — | — | — | — | — | — | — | V *d*-orbital of 1200 T (TrH distortion [1]) |
| $\chi_1$ | 1336 (0.76) | — | — | — | — | — | — | — | |
| $\chi_2$ | 1404 | — | — | — | — | — | 1389 | — | |
| High frequency ($F_h$) | | | | | | | | | |
| λ | 1550 (0.89, NTr) | 1605 (0.240, NTr) | 1700 | — | — | — | — | — | 1690 T (2×2×4 distortion [18]) |
| σ | 1705 | — | — | — | — | — | — | — | |
| ϖ | 1943 (0.69) | — | 1930 | — | — | — | — | — | 1831 T (circular FS at A [3]), 1940 T [18] |
| ϑ | 2037 (0.40) | — | — | — | — | — | 2085 (0.11) | — | Triangular orbit of 1967 T (TrH distortion [1]) |
| Λ | 2118 | 2135 (0.233, NTr) | — | — | — | — | 2167 (0.11), 2717 (0.13) | — | |
| Ultra-high frequency ($F_{uh}$) | | | | | | | | | |
| κ | 2753 (1.13, NTr) | — | — | — | — | — | — | — | 9329 T and 12846 T (undistorted FS [1]) |
| τ | 4025 (1.57, Tr) | — | — | — | — | — | — | — | |
| ∂ | 4827 | — | — | — | — | — | — | — | |
| ℵ | 5341 | — | — | — | — | — | — | — | |
| ℜ | 5996 (2.02) | — | — | — | — | — | — | — | |
| ℒ | 8014 | — | — | — | — | — | — | — | |
| 𝒳 | 9930 | — | — | — | — | — | — | — | |

## References in the Supplementary Materials